\documentstyle[prl,twocolumn,aps,epsf]{revtex}  
\begin{document}
\draft
\title{Magnetic Field Saturation in the Riga Dynamo Experiment}
\author{Agris Gailitis, Olgerts Lielausis, Ernests Platacis, 
Sergej Dement'ev,
Arnis Cifersons}
\address{Institute of Physics, Latvian University\\
LV-2169 Salaspils 1, Riga, Latvia}
\author{Gunter Gerbeth, Thomas Gundrum, Frank Stefani}
\address{Forschungszentrum Rossendorf\\
P.O. Box 510119, D-01314 Dresden, Germany}
\author{Michael Christen, Gotthard Will }
\address{Dresden University of Technology, Dept. Mech. Eng.\\
P.O. Box 01062, Dresden, Germany}

\date{\today}

\maketitle

\begin{abstract}

After  the  
dynamo experiment in November 1999 \cite{PRL1}
had shown magnetic field self-excitation   
in a spiraling liquid metal flow, 
in a second series of experiments 
emphasis was placed on the magnetic field saturation regime 
as the next principal step in the dynamo process. 
The dependence
of the strength of the magnetic field on the rotation rate
is  studied. Various features
of the saturated magnetic field 
are outlined and possible saturation 
mechanisms are discussed.
 
\end{abstract}

\pacs{PACS numbers: 47.65.+a, 52.65.Kj, 91.25.Cw}

\narrowtext

In the last decades, the 
theory of homogeneous dynamos has turned out unrivaled in 
explaining magnetic fields of planets, stars and galaxies. 
Enormous progress has been made in the numerical 
treatment of the so-called kinematic dynamo problem
where the velocity $\bf{v}$ of the fluid with electrical 
conductivity $\sigma$ is assumed to be given and
the behaviour of the magnetic field $\bf{B}$ 
is governed by the
induction equation 
\begin{eqnarray}
\frac{\partial {{\bf{B}}}}{\partial t}=\nabla 
\times ({\bf{v}} \times {\bf{B}})
+\frac{1}{\mu_0 \sigma} \Delta {\bf{B}} \; .
\end{eqnarray}
Above a critical value of the magnetic Reynolds number
$Rm=\mu_0 \sigma L v$ where $L$ and $v$ denote a typical 
length and velocity scale of the fluid, respectively, 
the obvious solution ${\bf{B}}=0$ of Eq. 1 may become unstable
and self-excitation of a magnetic field may occur.
The investigation of the saturation mechanism 
which limits the exponential growth of the magnetic field
is much more involved as it requires the simultaneous
solution of Eq. 1 and the Navier-Stokes equation 
 \begin{eqnarray}
\frac{\partial {{\bf{v}}}}{\partial t}+({\bf{v}} 
\cdot \nabla) {\bf{v}}=
- \frac{\nabla p}{\rho} + \frac{1}{\mu_0 \rho} 
(\nabla \times {\bf{B}}) 
\times {\bf{B}}+\nu \Delta  {\bf{v}}
\end{eqnarray}
where the back-reaction of the magnetic field on the velocity
is included. In Eq. 2, $\rho$ and $\nu$ denote
the density and the kinematic viscosity of the fluid, 
respectively. 

A number of recent 
computer simulations of the Earth's core has
led to impressive similarities between 
the computed and the observed magnetic field
behaviour including, most remarkably, 
field reversals \cite{GLRO}. 
The actual relevance of these simulations for real
dynamos is, however, not completely clear as 
 they are either 
working 
in parameter regions far from that of the Earth or 
have  resort to numerical trickery as, e.g.,  
the use of anisotropic 
hyperdiffusivities (for a recent overview, see \cite{BUSS}). 

Until very recently, a serious problem of the 
science of hydromagnetic 
dynamos was the lack of 
any possibility to verify numerical results
by experimental work. 
This situation 
changed on 11 November 1999 when in the first Riga dynamo 
experiment self-excitation of a 
slowly growing magnetic field eigenmode 
was observed
for the first time in an experimental liquid 
metal facility \cite{PRL1}.
On the background of an amplified external signal, 
the growth rate and the rotation frequency of the 
self-excited field were identified 
in good agreement with the numerical 
predictions based on kinematic dynamo theory.  
Due to some technical problems, the 
saturation regime was not reached in 
this first experiment. 
Soon after, Stieglitz and M\"uller studied
self-excitation 
and the saturation regime in another dynamo facility
in Karlsruhe  \cite{MUST}.

In the following we will represent the results of a second 
series of experiments at the Riga facility which were carried
out during 22-25 July 2000. It
was possible to work at considerably lower sodium temperatures 
and thus at higher electrical conductivities 
than in the November  experiment. This fact allowed to study 
in great detail 
the dynamo behaviour in the kinematic as well as in the 
saturation regime. Here, we will mainly focus on the 
results in the saturation regime as they go essentially
beyond the results of the November experiment.

The Riga dynamo facility consists of three concentric tubes 
of approximately 3 m length (Fig. 1). In the innermost tube of
25 cm diameter a spiraling flow of liquid sodium with 
a velocity up to 15 m/s is 
produced by a propeller. The sodium flows back 
in a second coaxial tube and stays at rest in a third
 outermost tube.
For more details of the experimental design see \cite{PRL1}
and references therein.

\begin{figure}
\begin{center}
\epsfxsize=5.5cm\epsfbox{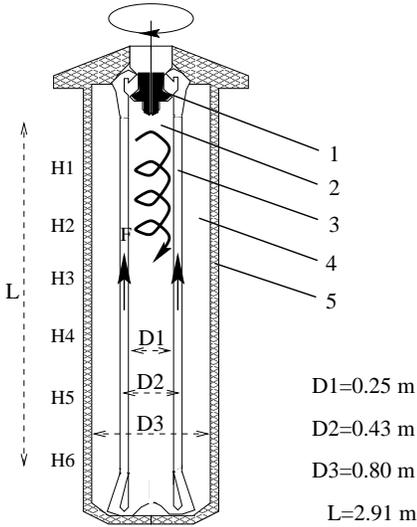}
\vspace{2mm}
\caption{The main part of the Riga dynamo facility: 
1 - Propeller, 2 - Helical
flow region, 3 - Back-flow region, 4 - Sodium at rest, 
5 - Thermal insulation
F - Position of the flux-gate sensor and the induction coil, 
H1...H6 - Positions of 
Hall sensors.}
\end{center}
\label{}
\end{figure}

As in November 1999, the $r$-components of the magnetic field 
outside the dynamo 
were measured 
by Hall sensors
at eight different places, six 
of which were aligned parallel to the 
axis. Within the dynamo, close to the innermost wall, 
the $z$-component of the magnetic field was  
measured by a sensitive magnetometer 
to study the starting phase of field generation
and the $r$-component by a less sensitive induction coil 
recording over all the generation time.

\begin{figure}
\begin{center}
\epsfxsize=7.5cm\epsfbox{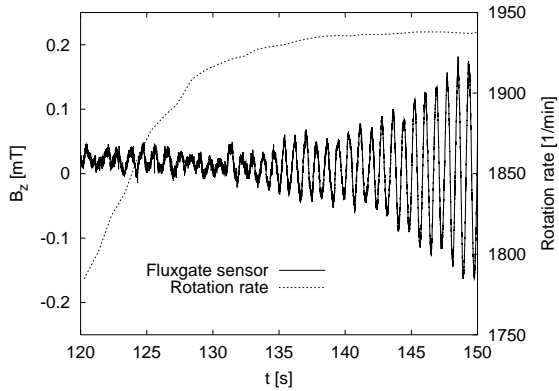}
\end{center}
\caption{The signal from the inner flux gate sensor (F in
Fig. 1) at the
beginning of magnetic field self-excitation together with 
the rotation rate.}
\end{figure} 

A typical run is documented in Figs. 2 and 3. Without any
external excitation, a rotating field starts to grow 
exponentially around $t=130$ s where the rotation rate 
has reached  1920 min$^{-1}$ (Fig. 2). 
Starting with this moment the rotation rate was kept practically 
constant for 80 s.
The whole magnetic 
field pattern rotates around the vertical axis in 
direction of the propeller rotation, but much slower. 
Hence, each sensor records an AC signal with a frequency
of the order 1 Hz.

At around $t=240$ s the 
exponential growth goes over into saturation
where the magnetic field continues to rotate 
with a certain frequency but stays essentially constant in its 
amplitude (Fig 3). Between $t=250$ s and $t=350$ s a number 
of different rotation rates has been studied, showing
a clear dependence of the saturation field level on the
rotation rate.

\begin{figure}
\begin{center}
\epsfxsize=7.5cm\epsfbox{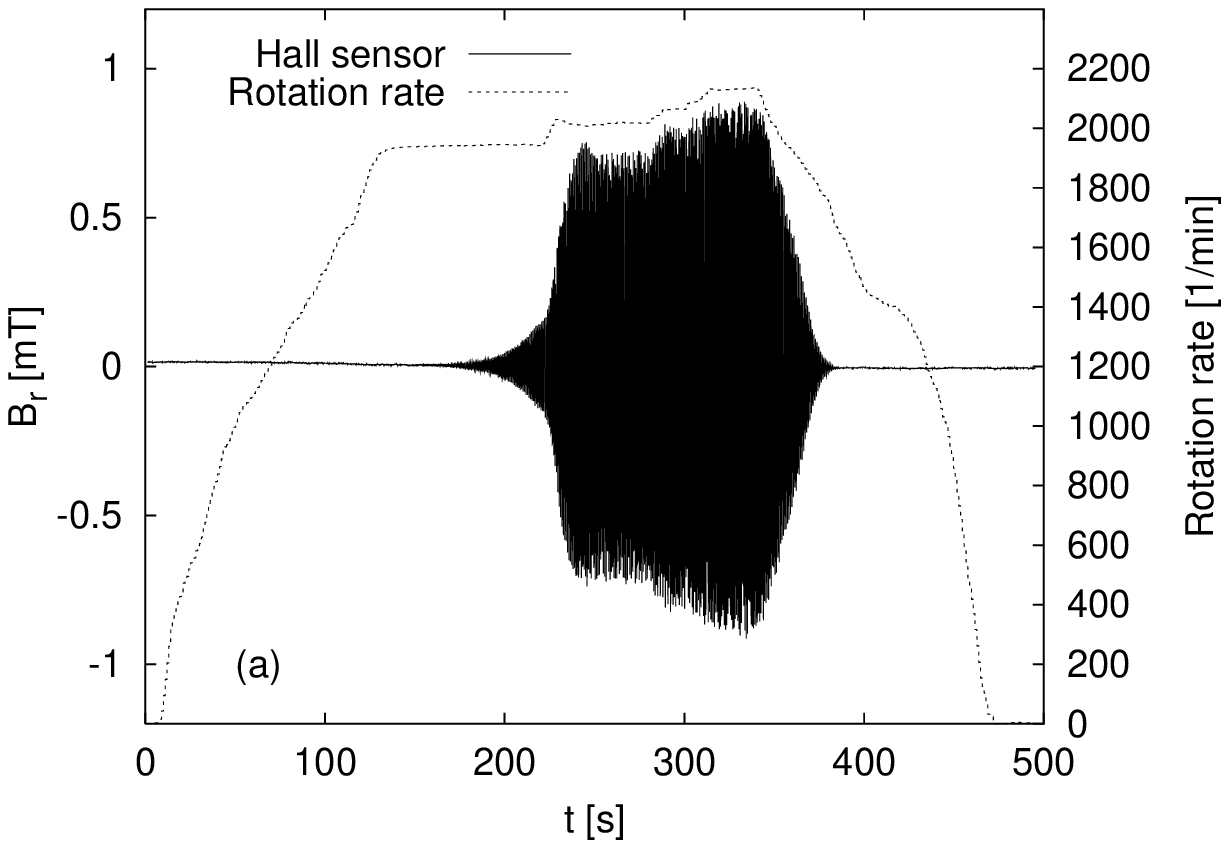}
\epsfxsize=7.5cm\epsfbox{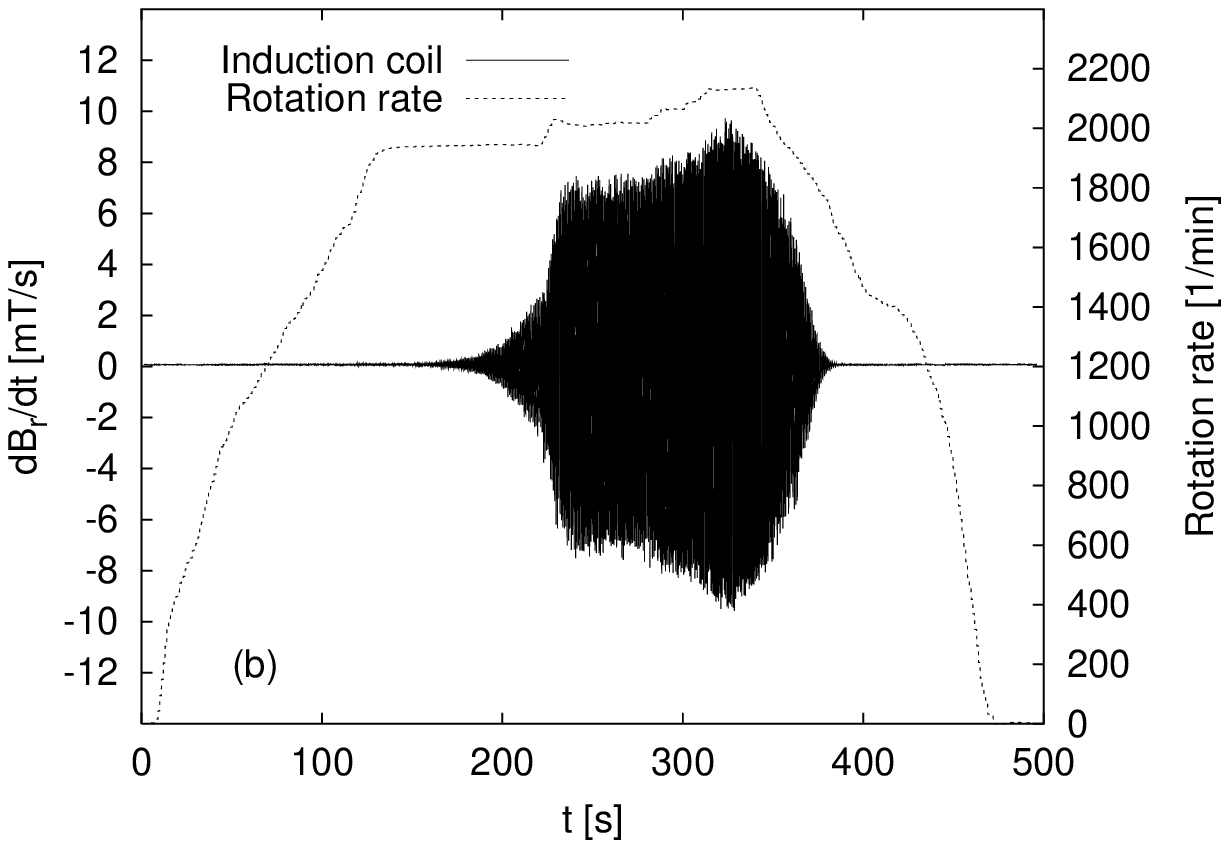}
\end{center}
\caption{One of the experimental runs with self-excitation and 
saturation: Signals 
measured at the Hall sensor H4 (a), and in the 
induction coil at position F (b).}
\end{figure}

From in total four experimental runs similar 
to that shown in Fig. 3, we compiled a number of
data concerning the dependence of the growth rates (Fig 4a), 
the frequencies (Fig. 4b), the power consumption
of the motors (Fig. 5), and the saturation levels (Fig. 6) 
on the rotation rate. 

The first observation to be made in Fig. 4 is a satisfactory 
confirmation of the numerical predictions, in particular with
respect to the slope of the curves and the temperature 
dependence. There is a parallel shift 
of approximately 10 per cent with respect to the  
growth rate and of approximately 5 per cent 
with respect to the frequency. The shift of the 
growth rates might be partly explained by the fact that
the steel walls (with their lower electrical conductivity) 
were not taken 
into account in the 2D code underlying 
the given numerical predictions. 
From
one-dimensional calculations it is known that the general effect 
of these walls consists in an increase of the critical $Rm$ by 
about 8 
per cent which is also the order of the said parallel shift.
In total, there is now a quite 
coherent picture of the kinematic regime in which 
the results from the November experiment and the recent results 
fit together.

\begin{figure}
\begin{center}
\epsfxsize=7.5cm\epsfbox{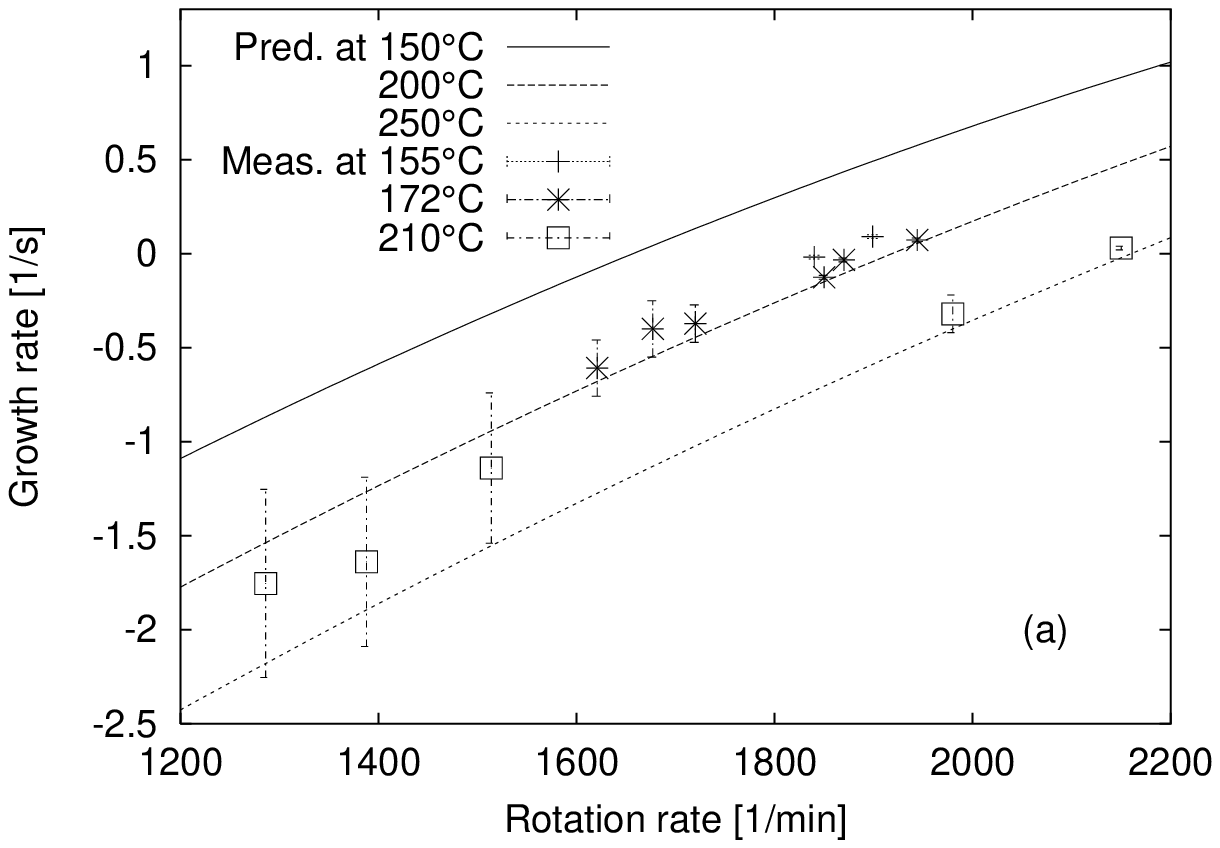}
\epsfxsize=7.5cm\epsfbox{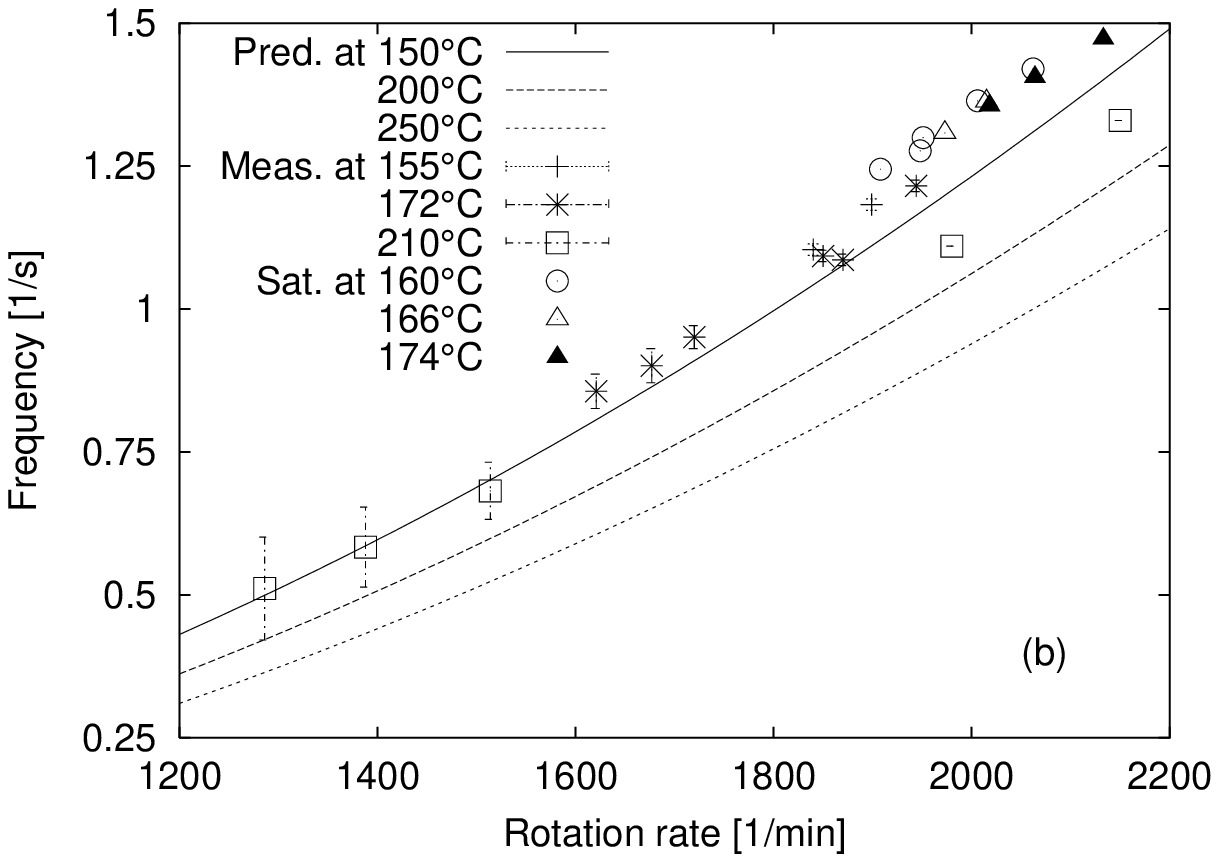}

\end{center}
\caption{Growth rates (a) and frequencies (b) for different 
rotation rates and temperatures in comparison with the numerical 
prediction. For the sake of clarity, 
data measured at close temperatures are grouped together. 
The shown temperatures are to 
be understood as $\pm$ 3$^{\circ}$C.
The two rightmost points at 210$^{\circ}$C are taken over from the 
November experiment.
The larger error bars at low rotation rates
are due to the fact that for a quickly decaying signal
the frequency and the decay rate are much harder to determine than
for a slow decaying or an increasing signal. In (b) the frequencies
in the saturation regime are also given, whereas the corresponding
growth rates are (by definition) zero.}
\end{figure}

As for the saturation regime there is an interesting 
observation to be made in Fig. 4b. Whereas in this
regime the growth rates are by definition zero, 
the frequencies continue to 
increase with the   rotation rate. Obviously, 
the frequency values are even a bit higher than what would
 be expected from a simple extrapolation of the data in the 
kinematic regime. This fact indicates 
that there is not only an overall brake of the sodium flow 
by the Lorentz forces but that the spatial structure of the
flow is changed, too. We will come back to this point later
on. 

Concerning the motor power increase in the saturation regime 
in Fig. 5 we show for one experimental run 
the motor power  consumption as a function of 
the rotation rate $\Omega$. The points in Fig 5 correspond
to those time intervals where the rotation rate was 
kept constant. 
The shown data in the field-free 
regime are best fitted with an $\Omega^3$ curve which is
also expected from simple 
hydraulic arguments.
As for the four rightmost points in the saturation regime, there
is evidently an increase of the power consumption which must be
explained in terms of the back-reaction of the magnetic field.
In order to estimate this additional power the Ohmic losses
$P_{Ohm}=\int j^2/\sigma\;  dV$ have to be computed. 
Using the magnetic field structure as it results from  the
kinematic code and fixing its strength to the
measured field strength
we get  Ohmic losses of approximately 
$10$ kW which is a good approximation
of the observed power consumption 
increase.

\begin{figure}
\begin{center}
\epsfxsize=7.5cm\epsfbox{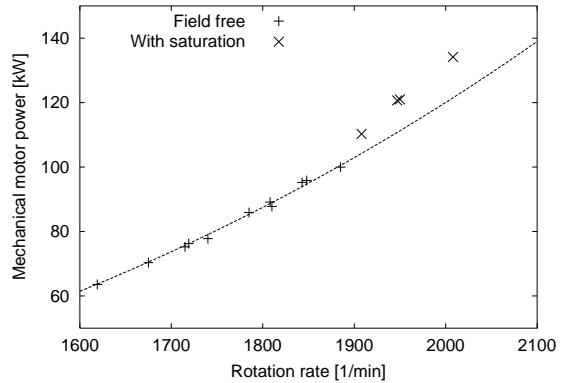}
\end{center}
\caption{Motor Power below and in the saturation regime }
\end{figure}

However, as already mentioned in connection with the
observed frequency behaviour, the power increase due to  
magnetic forces seems to be not the whole story. Another 
feature of the  field might help to identify 
a further saturation mechanism. In Fig. 6 we have plotted 
four  saturation levels of the radial magnetic field 
component measured at 
different positions in
dependence of the temperature corrected rotation rate $\Omega_c$.
Three of the measurements where carried out at the
Hall sensors H2, H4, H6 (see Fig. 1). The fourth 
field is inferred from the 
induced voltage (divided by the frequency) in the 
induction coil
at position F. The magnitude of the radial component
reaches 7 mT. From numerical simulations of the
kinematic regime it is known that
the axial component at the given 
position is by a factor five larger than the radial component. 
Hence a value of about 30 mT 
seems to be a reasonable estimate for this 
$z$-component  inside the dynamo.
The correction of the rotation rates 
has been done the following way: at first note
that at a temperature of 157$^{\circ}$C and at 
a rotation rate of 1840 min$^{-1}$ 
an extremely slowly decaying
mode was observed in one run which will serve as
a definition of a marginal state. Saturation was 
observed, however, at quite different temperatures. 
Therefore, 
the measured rotation rate $\Omega$ at the temperature $T$ 
was first corrected corresponding
to $\Omega_{c}(T)=\sigma(T)/\sigma(157^{\circ}C) \, \Omega(T)$. 
By virtue of this definition, the corrected 
rotation rate $\Omega_{c}$ is proportional to the
magnetic Reynolds number $R_m$.
The four different saturation levels in Fig. 6 are fitted 
by some curves proportional to 
($\Omega_{c}$ - 1840 min$^{-1}$)$^{\beta}$ 
(this is no pre-justice for a corresponding behaviour 
close to $\Omega=$1840 min$^{-1}$; a slightly 
sub-critical behaviour 
close to the marginal point cannot be
excluded by our data). 
The different exponents give clear evidence for a 
remarkable redistribution of the magnetic energy towards
the propeller region. 

In trying to explain the two features of the magnetic field 
which go beyond an  overall braking of the flow by  
the magnetic 
pressure, namely, the relatively high frequency and the 
redistribution of magnetic energy towards 
the propeller region, one 
can utilize  numerical results for those flows with 
an azimuthal velocity component which is  decaying
along the z-axis.
To begin with, note that 
the self-excited magnetic field with 
its azimuthal dependence $\exp(im\phi)$ with $m=1$ gives 
rise to 
Lorentz forces with $m=0$ and $m=2$. Of the former, 
which will 
be the only one of interest in the following, there 
are components which can be absorbed into  pressure 
derivatives in $z$ and $r$ direction. The $\varphi$ component, 
however, remains. 
The perturbation $\delta v_{\phi}$ of the azimuthal
velocity component arising from the Lorentz forces
can be estimated from Eq. 2 to be in the order of
1 m/s, if one takes into account the measured magnetic field.
It is interesting to observe that the numerical
simulation of the influence 
of the azimuthal velocity changes along the $z$-axis for the
kinematic dynamo \cite{STEF} results in similar magnetic 
field redistributions as shown for the
saturation in Fig. 6. Hence,
this decay 
might indeed be relevant and it is therefore 
the  favorite candidate to account 
for the observed features
of the magnetic field in the saturation regime.

\begin{figure}
\begin{center}
\epsfxsize=7.5cm\epsfbox{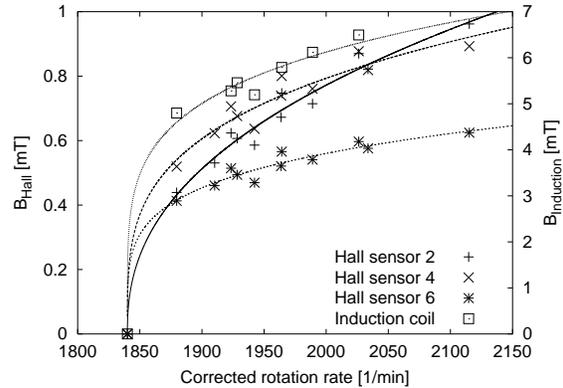}
\end{center}
\caption{Measured magnetic field levels in the saturation regime.
The corrected rotation rate $\Omega_{c}$ 
is related to the measured rotation rate
$\Omega$ via 
$\Omega_{c}(T)=\sigma(T)/\sigma(157^{\circ}C) \, \Omega$.
The lines are interpolating curves of the form 
($\Omega_c-$1840 min$^{-1})^{\beta}$ with $\beta=0.417$ for H2, 
$\beta=0.275$ for H4, $\beta=0.216$ for H6, $\beta=0.205$ for
the induction coil.}
\end{figure}

The study of the saturation regime will be continued by
means of numerical simulations in tight connection with a
sort of inverse dynamo approach based on 
the measured data and, hopefully, with direct
velocity measurements inside the dynamo facility.

We thank the Latvian Science Council for support under grant 96.0276,
the Latvian Government and International Science Foundation for support
under joint grant LJD100, the International Science Foundation for support
under grant LFD000 and
Deutsche Forschungsgemeinschaft for support under INK
18/B1-1.

% that you will use with \ref{} command in the braces of the \label{}
% command.
% that you will use with \ref{} command in the braces of the \label{}
% command.
% \begin{table}
% \caption{}
% \label{}
% \end{tabular}
% \end{table}

\end{document}